\documentclass{aa} 
\usepackage{graphicx}
\usepackage{txfonts}
\begin{document} %
\title{Discovery of multiple ultra-luminous X-ray sources in the galaxy
KUG 0214-057\thanks{Based on observations obtained with XMM-Newton, an
ESA science
mission
with instruments and contributions directly funded by ESA Member States
and NASA.}}

\author{M G Watson\inst{1}
\and
T P Roberts\inst{1}
\and 
M Akiyama\inst{2}
\and
Y Ueda\inst{3}}

\offprints{MGW (\email{mgw@star.le.ac.uk})}

\institute{Department of Physics \& Astronomy, University of Leicester,
Leicester, LE1 7RH, UK
\and
Subaru Telescope, National Astronomical Observatory of Japan, 650
North A'ohoku Place, Hilo, HI 96720, USA
\and
Institute of Space and Astronautical Science, Sagamihara,
Kanagawa
229-8510, Japan}
\date{Received 21 December 2004 / Accepted 06 April 2004 }

\abstract{We report the serendipitous discovery of several
unresolved X-ray sources lying in the prominent spiral arms of the galaxy
\object{KUG 0214-057} in {\it XMM-Newton\/} observations. The location of
these
X-ray sources strongly suggests that at least three, and possibly four, of
these may be physically related to the galaxy. The luminosity of each
of these sources at the distance of KUG 0214-057 is $> 5 \times
10^{39}\rm\ erg\ s^{-1}$ (0.3-10 keV), making each a strong candidate
ultraluminous
X-ray source (ULX). Using the ULX objects as a metric implies that this
relatively low-mass
galaxy may be experiencing rather intense starburst activity. The
serendipitous discovery of these ULX objects suggests that such objects
are not a negligible component of the overall extragalactic X-ray source
population.

\keywords{X-rays: galaxies -- Galaxies: starburst}
   }

\maketitle
%
%________________________________________________________________

\section{Introduction} 

Ultraluminous X-ray sources (ULXs) are the second most luminous class
of discrete X-ray sources associated with external galaxies, only
surpassed in luminosity terms by active galactic nuclei (AGN).
However, unlike AGN, they can be located anywhere within the confines
of a galaxy.  ULXs are observationally defined (eg. Makishima et
al. \cite{mak00}) as discrete X-ray
sources displaying a luminosity in excess of $10^{39} \rm~erg~s^{-1}$,
equal to the Eddington luminosity for accretion onto a $\sim
7$ M$_{\odot}$ black hole.  Hence the high apparent luminosities of
many ULXs, particularly those exceeding $10^{40} \rm~erg~s^{-1}$, have
led to a debate on their underlying nature (see Miller \& Colbert \cite{mil04}
for a recent review).  In particular, this debate focuses on whether
any ULXs are powered by accretion onto ``intermediate-mass'' black
holes (IMBH, with M$_{\rm BH} \sim 10^2 - 10^5$ M$_{\odot}$;
e.g. Colbert \& Mushotzky \cite{col99}; Miller et al. \cite{mil03}), or whether
instead accretion onto a stellar-mass black hole is responsible
(M$_{\rm BH} < 20$ M$_{\odot}$, as seen for Galactic black holes),
where the high luminosities originate either in truly super-Eddington
radiation from inhomogeneous accretion discs (Begelman \cite{beg02}), or
through anisotropic radiation patterns and/or relativistic beaming
(e.g. King et al. \cite{kin01}; Reynolds et al. \cite{rey97}).

Perhaps the most interesting result to emerge from {\it Chandra\/}
observations of starburst galaxies has been the detection of significant
populations of ULXs, many of which have extreme luminosities
i.e. close to or in excess of $10^{40} \rm~erg~s^{-1}$, in the most
active star forming regions of these galaxies (e.g. NGC 3256, Lira et
al. \cite{lir02}; the Antennae, Fabbiano et al. \cite{fab01}; the Cartwheel, Gao et
al. \cite{gao03}).  Whilst it is possible that some of these ULXs may be
powered by IMBHs formed in young, dense star clusters (as now seems
likely for the extreme ULX in M82; Portegies Zwart et al. \cite{por04}), this
formation route is unlikely for the majority of ULXs in these systems,
which are instead likely to be ``ordinary'' high-mass X-ray binaries
powered by one or more of the mechanisms outlined above (King \cite{kin04}).
Regardless of their origin, it is now becoming apparent that the
presence (and integrated luminosity) of multiple ULXs may be a new
gauge of high rates of star formation activity (Grimm, Gilfanov \&
Sunyaev \cite{gri03}; Persic et al. \cite{per04}), and indeed the numbers of galaxies
containing ULXs has been demonstrated to rise proportionally with the
star formation rate out to $z\approx 0.1$ (Hornschmeier et al. \cite{hor04}).

Most of the ULXs known have been discovered in pointed observations of
specific galaxies, ie. essentially in surveys of external galaxy populations,
although their discovery has often been fortuitous.  In this paper we report
the serendipitous discovery of several unresolved X-ray sources lying in the
prominent spiral arms of the galaxy \object{KUG 0214-057} in {\it
XMM-Newton\/}
observations of this region of sky, made as part of the Subaru {\it
XMM-Newton\/} Deep Survey (SXDS). The location of these X-ray sources strongly
suggests that at least three, and possibly four, of these may be physically related to the galaxy
as opposed to foreground or background objects. The luminosity of each of these
sources, if at the distance of \object{KUG 0214-057}, is $> 5 \times
10^{39}\rm\ erg\
s^{-1}$, placing all three towards the high luminosity end of ULX regime.
These results thus correspond to the discovery of ULXs as part of an unbiased
extragalactic survey.

\begin{figure*}
\centering
\includegraphics[width=18cm]{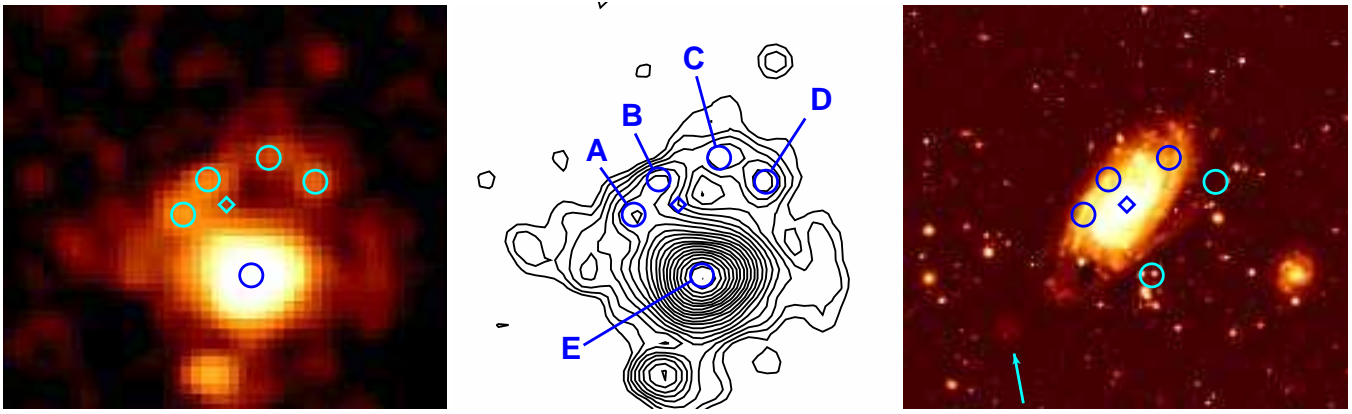}
\caption{(a) \& (b) ({\sl left and centre panels}):  The soft band (0.5-2
keV) {\it XMM-Newton\/} X-ray
image centred on \object{KUG 0214-057}. Data shown is for the 3 EPIC cameras
combined in both colour and contour representations. The
circles, plotted with nominal 5 arcsec.  radius, mark the 5 sources
discussed in the text, labelled with source identifiers as in Table 1.
Note that source E is strongly saturated in the image shown in (a).
The location of the nucleus of \object{KUG 0214-057} is marked with a
diamond in each panel. The X-ray image has been
smoothed with a Gaussian with $\sigma =5$ arcsec.  (c) ({\sl right-hand
panel}):
Optical R-band
image of region containing \object{KUG 0214-057} obtained with Subaru
SuprimeCam.
The smaller galaxy to the E is \object{APMUKS(BJ) B021449.37-054240.9};
the nebulosity discussed in the text is the faint feature to the SE marked
with an arrow.
Each panel has dimensions $\sim$ 3 x 3 arcmin.}
\end{figure*}

\begin{table*} 

\begin{minipage}[t]{\textwidth}
\caption{Summary
properties of the X-ray sources in the region of \object{KUG 0214-057}}
\centering
\begin{tabular}{cccrrrrrrrl} \hline\hline Source\footnote{IAU designations of 
sources are as follows:
{\bf A}~XMMU J021729.6-052855;
{\bf B}~XMMU J021719.9-052840; XMMU J021718.1-052831;
{\bf D}~XMMU J021716.8-052841;
{\bf E}~XMMU J021718.6-052921} & RA & Dec &
\multicolumn{1}{c}{count rate\footnote{X-ray count rate in 0.3-10 keV
band expressed in EPIC
pn count s$^{-1}$}} & \multicolumn{1}{c}{HR1\footnote{Hardness ratios HR1,
HR2 \&HR3 are defined in the caption to Fig.3.}}
&\multicolumn{1}{c}{HR2} &\multicolumn{1}{c}{HR3} & $\log_{10}
f_X$\footnote{X-ray flux in 0.3-10 keV band [$\rm erg\ cm^{-2}\
s^{-1}$] assuming a nominal power-law spectrum with $\Gamma =1.7$ and
absorption column density $N_{H}=2.5\times 10^{20}\rm cm^{-2}$}&
\multicolumn{1}{c}{$z$\footnote{Assumed redshift (see text)}}&$\log_{10}
L_X$\footnote{X-ray luminosity in 0.3-10 keV band [$\rm erg\
s^{-1}$]}\\
name       & \multicolumn{2}{c}{J2000.0}  &
       \multicolumn{1}{c}{[count
       s$^{-1}$]} &  &  &  & & &  \\
       
       \hline
A&02 17 20.60&-05 28 55.1&
0.00499\ $\pm$0.00043&0.83\ $\pm$0.07&-0.59\ $\pm$0.07&-0.59\
$\pm$0.29&-13.87&0.018& 39.98\\
B&02 17 19.88&-05 28 39.8&
0.00377\ $\pm$0.00039&0.81\ $\pm$0.08&-0.59\ $\pm$0.07&-0.41\
$\pm$0.34&-13.99&0.018& 39.86\\
C&02 17 18.13&-05 28 30.5&0.00370\ $\pm$0.00044&0.63\ $\pm$0.09&-0.68\ $\pm$0.08&-0.25\
$\pm$0.38&-14.00&0.018& 39.85\\
D&02 17 16.77&-05 28 40.8&
0.00245\ $\pm$0.00036&1.00\ $\pm$0.04&-0.50\ $\pm$0.12&-0.44\
$\pm$0.37&-14.17&0.018&39.67\\
E&02 17 18.62&-05 29 21.3&
0.12428\ $\pm$0.00131&0.57\ $\pm$0.01&-0.73\ $\pm$0.01&-0.42\
$\pm$0.02&-12.47&0.634& 44.76\\
& \\
\hline
\end{tabular}
\end{minipage}
\end{table*}

\section{Observations and data analysis}
The X-ray observations constituting the SXDS comprise a mosaic of 7 partially
overlapping pointings with {\it XMM-Newton\/} covering $\approx 1.2$ sq.deg.
with exposure times of 100 ksec (central field) and 50 ksec (flanking fields).
The results presented here are drawn from the complete X-ray source catalogue
for the SXDS region which contains $>$ 1000 objects (Ueda et al., 2005, in
preparation). The SXDS X-ray catalogue was produced standard {\it
XMM-Newton\/} SAS tools and
utilises an approach very similar to the standard pipeline processing
(Watson et al. \cite{wat01}\footnote{see also the documentation for the 1XMM 
catalogue which uses a very similar procedure:
http://xmmssc-www.star.le.ac.uk/}).
Small improvements were employed, eg. to optimise the background estimation and to
ensure closely-spaced sources are correctly resolved.

Here we concentrate on the ``SDS-6'' field, which was observed for a total
exposure time $\approx 50$ ksec. The SDS-6 field (OBSID: 0112370701,
observations made on 2002-08-08/2002-08-09) contains a small region
notable in having a relatively high local source density. Our attention was
first drawn to this source complex because of the rather striking pattern: a
single bright source surrounded by a partial ring of fainter objects (Fig.1a,b).
As is
discussed later, there is no reason to suppose that the bright X-ray source
and fainter objects are physically related, but it is very likely that several
of the faint X-ray sources are physically related to a galaxy they are
spatially coincident with.

Table 1 presents the basic parameters of the five X-ray sources detected by
{\it XMM-Newton\/} in this region of sky taken from the complete X-ray source
catalogue for the SXDS (Ueda et al., 2005, in preparation). The catalogue is
based on an analysis of the summed EPIC images, ie. the EPIC MOS1, MOS2 and
pn data combined; all the results discussed in this paper relate to the
combined EPIC data.
Source
parameters are those derived from the SAS task {\tt emldetect} which
determines the source parameters by fitting the instrumental point spread
function (PSF) at candidate source positions as part of the standard source
detection and parameterisation approach. All 5
sources reported in Table 1 are detected at relatively high significance (likelihoods
$\ge 30$ in the {\bf combined} EPIC data; sources C \& D have the lowest likelihoods) and are
confirmed by visual examination of the EPIC pn images and in the combined
EPIC MOS images (the lowest significance sources are only marginally
visible in the individual EPIC MOS images).
The count rates
reported are background-subtracted values corrected for vignetting and
correspond to the count rates integrated over the entire instrument PSF.

Fig. 1a,b shows the
{\it XMM-Newton\/} X-ray image with the source locations overlaid. By
cross-correlating the full set of X-ray sources detected in the SXDS region
with the Subaru optical object catalogue, the relative astrometry between the
Subaru and {\it XMM-Newton\/} astrometric frames has been established to
$<1$ arcsec.  The positions quoted in Table 1 have been corrected for small
systematic shifts in the {\it XMM-Newton\/} astrometric frame, ie essentially
to the reference frame defined for the Subaru imaging which has an absolute
astrometric accuracy better than 0.5 arcsec. The {\sl statistical} error
(1$\sigma$) on the X-ray source positions is $<0.5$ arcsec. for the brightest
source (E) and ranges from 1.2 to 1.8 arcsec. for the four fainter
sources listed. We adopt a 5 arcsec radius for the X-ray source positional
uncertainty.

\begin{figure*}
\centering
\includegraphics[height=5.5cm]{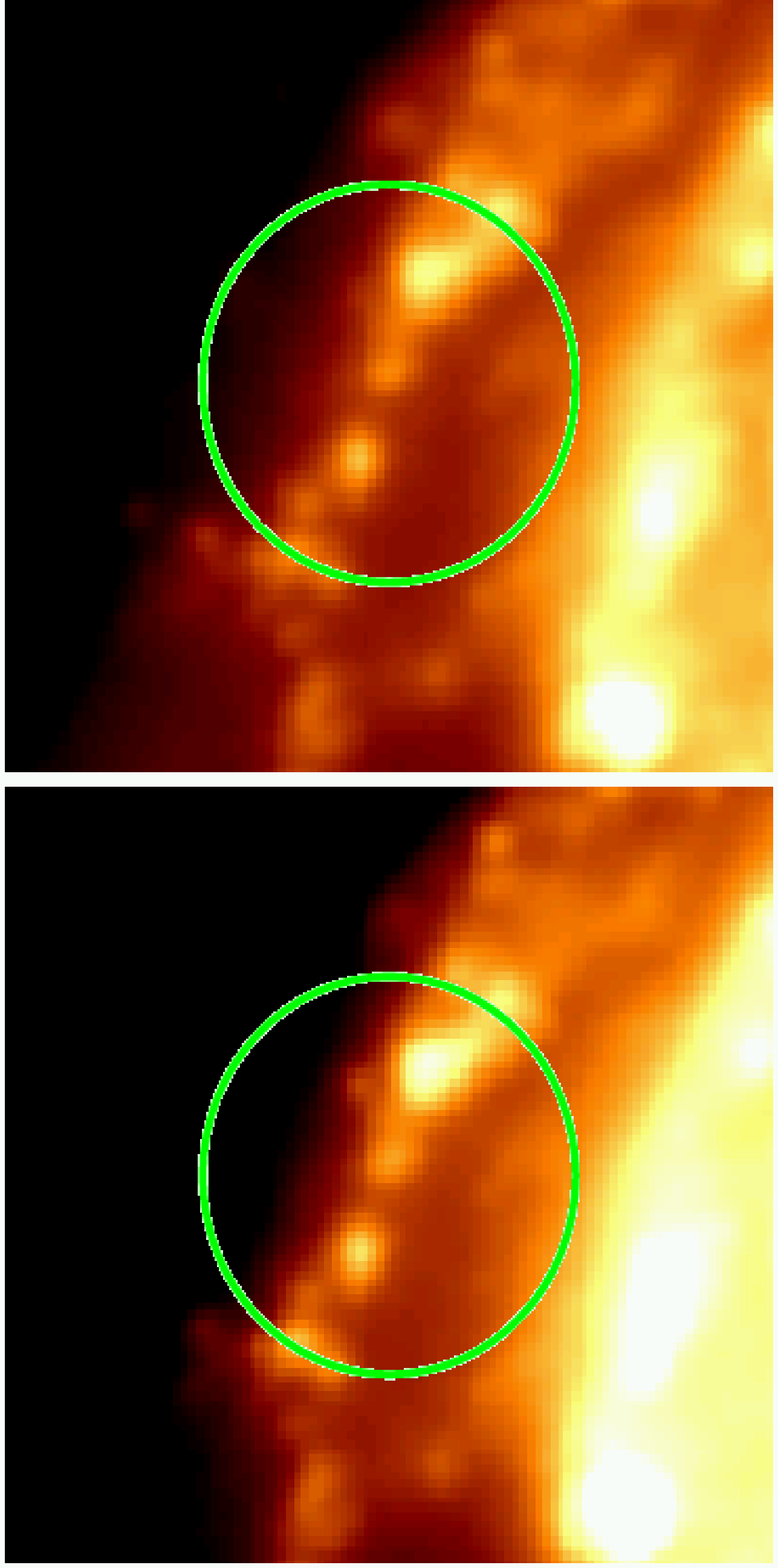}
\includegraphics[height=5.5cm]{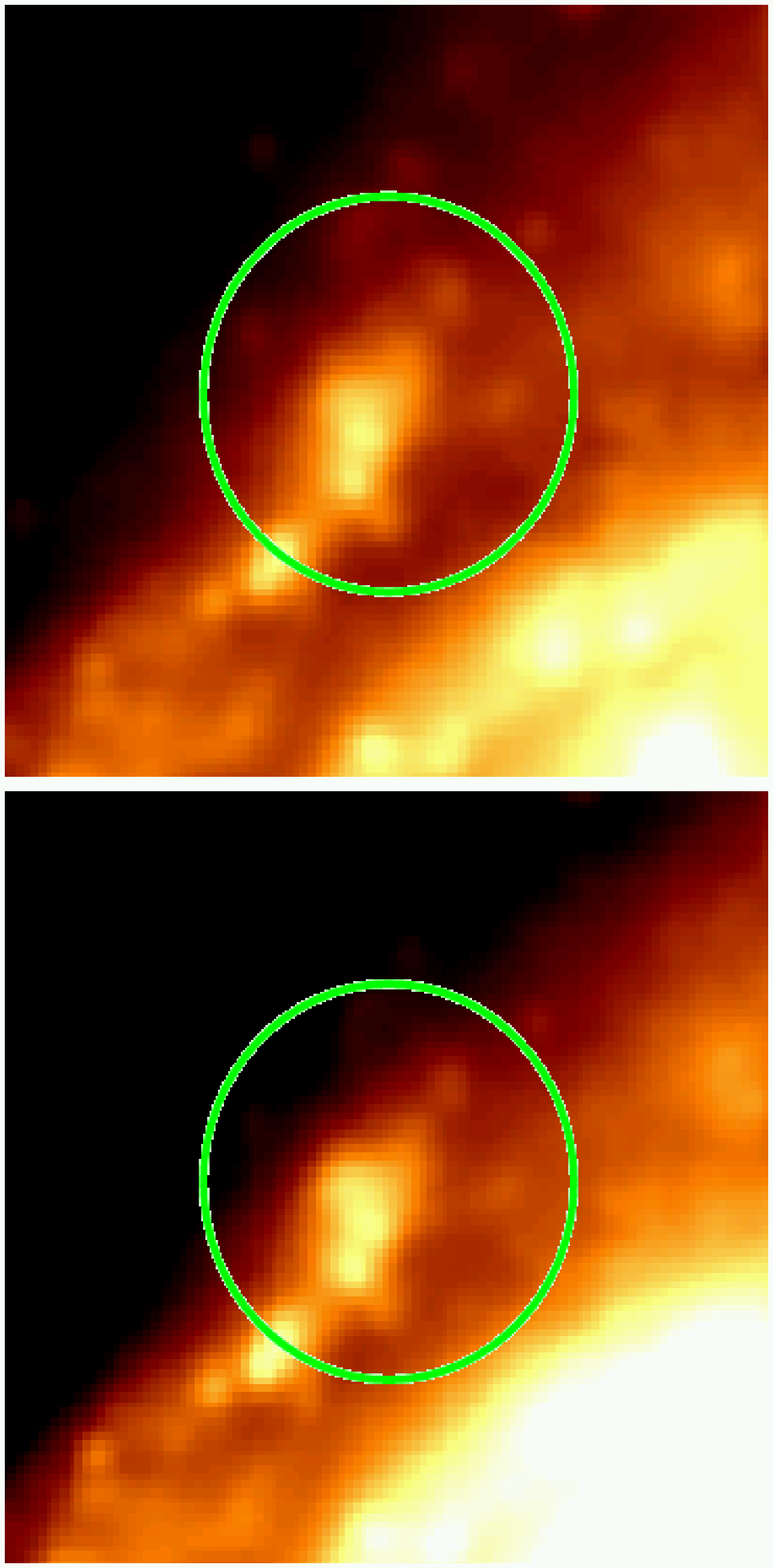}
\includegraphics[height=5.5cm]{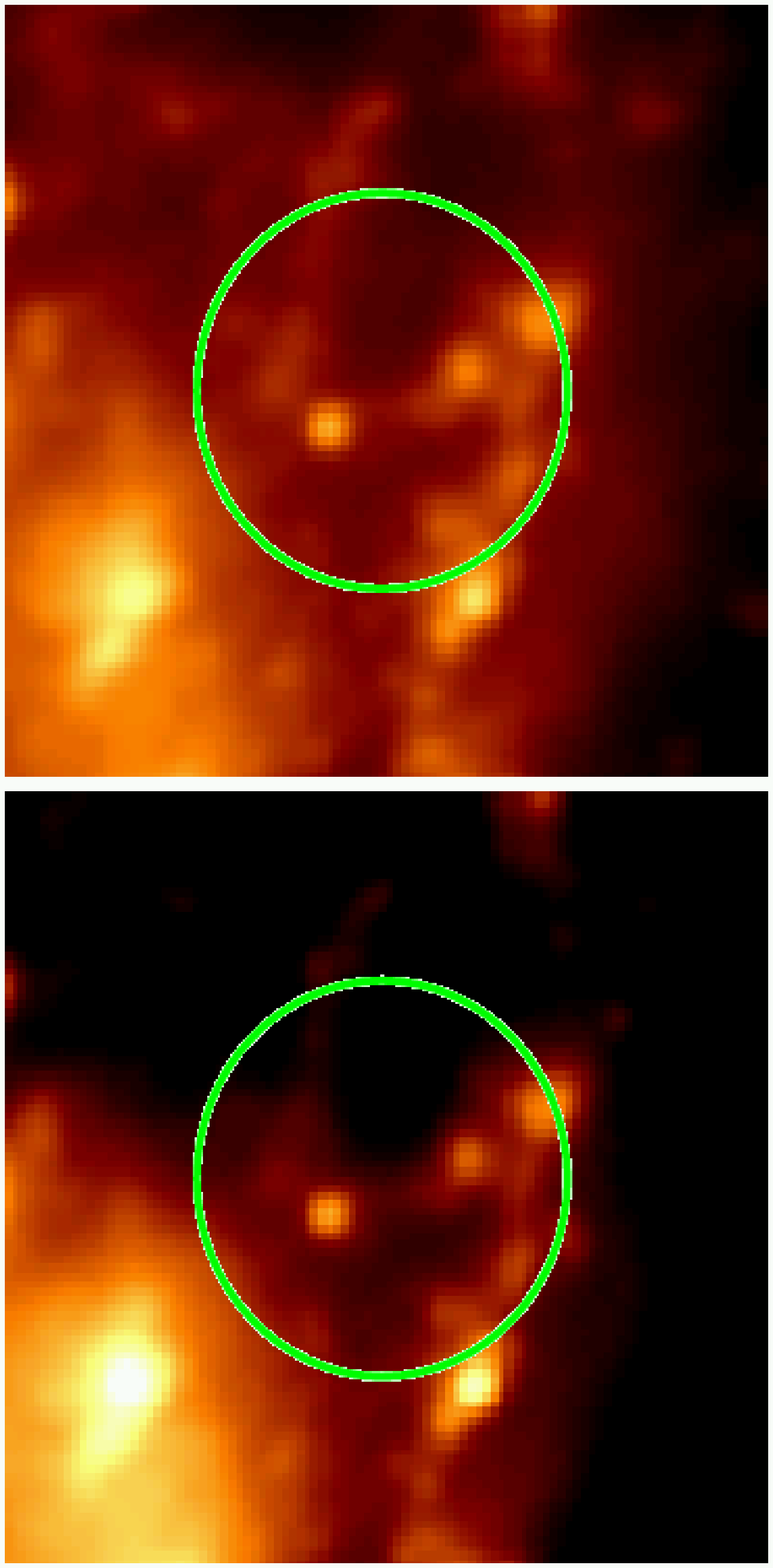}
\includegraphics[height=5.5cm]{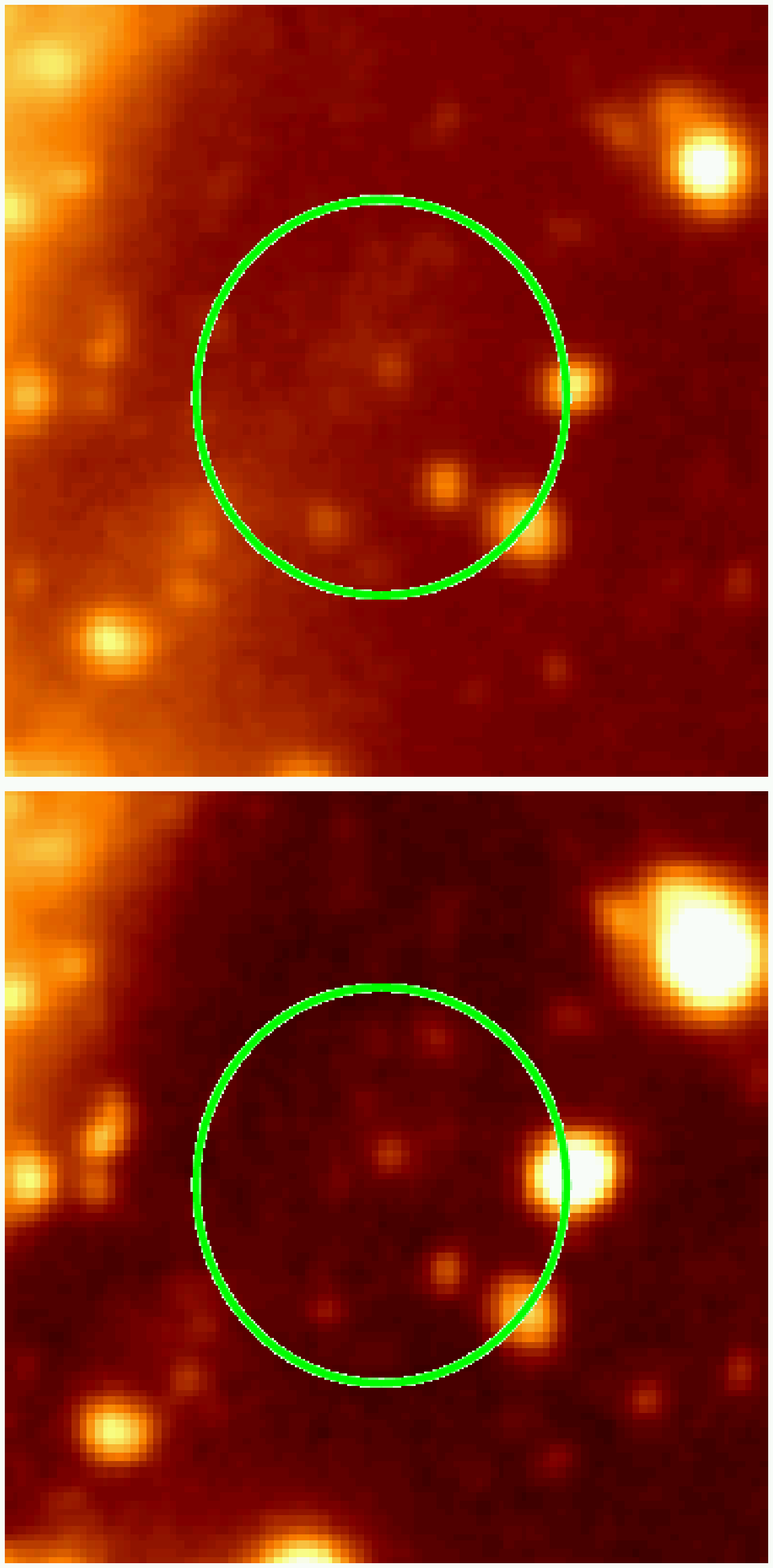}
\includegraphics[height=5.5cm]{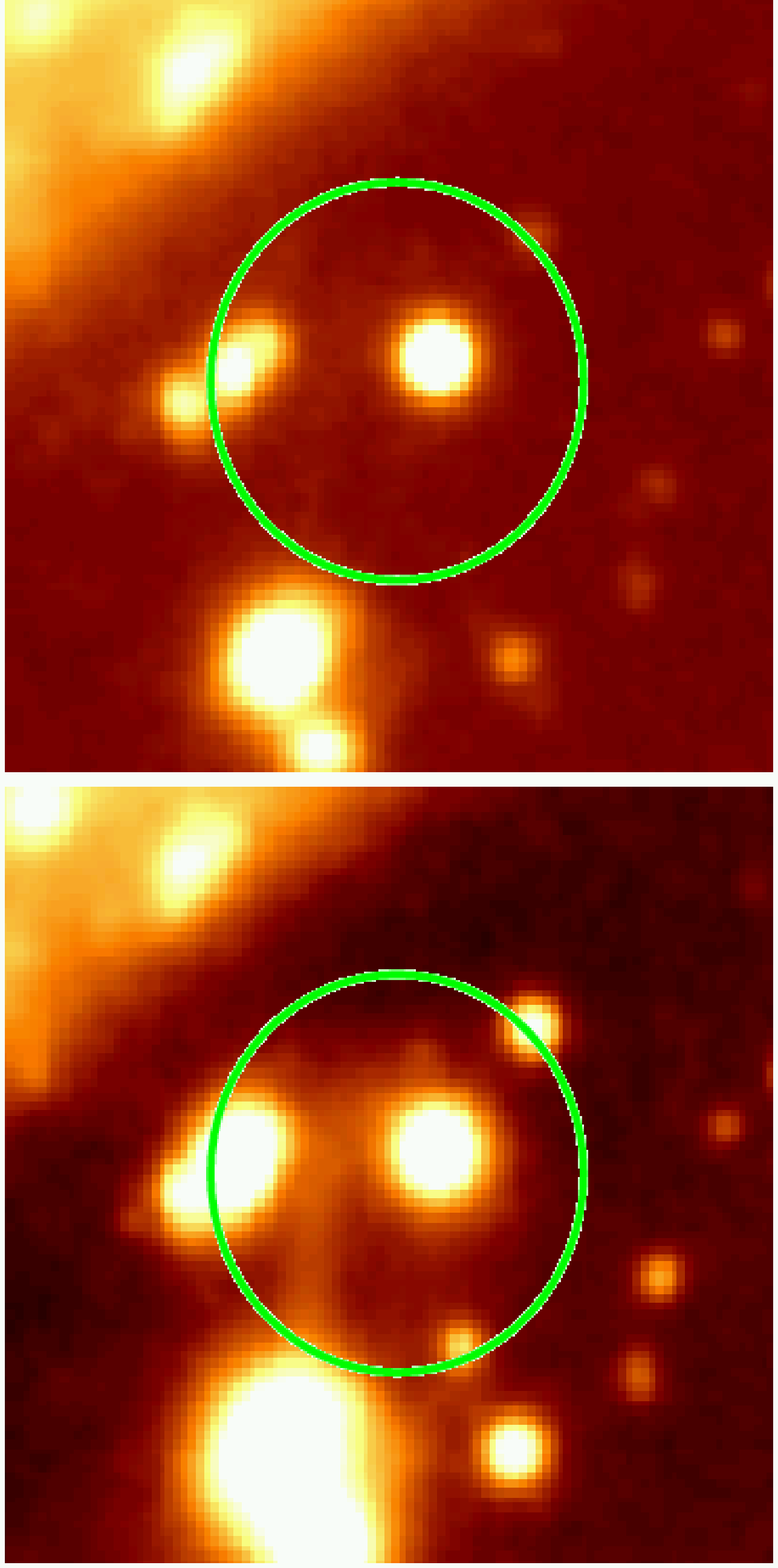}
\caption{Optical B-band (top row) and R-band (bottom row) finding charts for
the four ULX candidate
X-ray sources (A, B, C, D from left to right) and the background AGN (E).
The finding charts are
based on the Subaru SuprimeCam SXDS imaging. X-ray error-circles,
plotted with nominal 5 arcsec.  radius, are shown in green. Note that
the two right hand images are displayed with different colour table to
enhance the visibility of faint objects.}

\end{figure*}

The location of the X-ray sources on the optical sky is shown in
Fig. 1c.  The optical image shown is the B-band image obtained as part
of the Subaru SuprimeCam observations of the SXDS. The full SXDS
optical imaging consists of five colours (BVRi'z') and extends to very
faint limits (eg.  B$\approx 27.5$ mag.; AB; 5 $\sigma$) with
excellent seeing ($\approx 0.8$ arcsec.) (Furusawa et al. 2005, in preparation).

\section{Results and Discussion}
\subsection{Location and nature of the X-ray sources}
As can be seen in Fig. 1c, all five\footnote{ Note: Fig. 1a,b shows a
sixth bright X-ray source at the lower edge of the image.  This is
identified with an AGN at z=0.38 and is hence not connected with the
galaxy.}  X-ray sources lie in the outer regions of the 15 mag. galaxy
\object{KUG 0214-057} (Miyauchi-Isobe \& Mahehara \cite{miy98}; \object{KUG
0214-057} =
MCG -01-06-080/PGC 008726).  
\object{KUG 0214-057} is a barred
spiral galaxy at z=0.018 (corresponding to a distance $D\approx
75$ Mpc), lying in the  galaxy group NOG 151 with NGC 881 and NGC
883 (Giuricin et al. \cite{giu00}). Three of these sources (A, B and
C) are located in the outer spiral arms of the galaxy, whilst the
other two (E, D) are outside the visible optical envelope.

The brightest source, E, has been identified with an AGN at z=0.634 as
part of the ongoing optical follow-up of the SXDS X-ray sources (Akiyama et
al., 2005, in preparation); the optical counterpart is the brightest object
within the X-ray error-circle shown in Fig.2. As \object{KUG 0214-057} is
at z=0.018, the X-ray source E is clearly an unrelated background
object.\footnote{We note that the R-band finding chart for source E (Fig.2) shows
evidence for diffuse emission, including an unusual ``bridge'' feature, which 
may indicate the AGN galaxy is part of an interacting group.}

The location of the remaining four objects is most plausibly explained by them
being located within the galaxy itself. The evidence is arguably strongest for
sources A, B and C which are embedded in regions of the galaxy
rich in knots of nebulosity which, on the basis of their broad-band optical
colours, are very likely to be star-formation regions (see Fig.2 and section
3.3). The evidence is weaker for D which lies somewhat outside the
visible optical envelope of the galaxy, although not far from the end of the
apparently disrupted spiral arm to the W. If this is correct, the broad band
X-ray luminosities of these sources range from $\sim 5\times 10^{39}\rm\ erg\
s^{-1}$ to $\sim 10^{40}\rm\ erg\ s^{-1}$, making them ULXs with luminosities
towards the high end of those observed in external galaxies.

These X-ray sources might alternatively be foreground or background objects
unrelated to the galaxy. Based on the well-established high Galactic 
latitude X-ray
source $\log$ N -- $\log S$ relationship, the {\sl a priori} chance
probability of finding three or four X-ray sources at these fluxes within 1
arcmin.  radius of the galaxy is $< 1$\% and the probability of finding them
in the smaller region defined by the spiral arms even lower. More than 80\% of
faint high latitude X-ray sources are AGN, making them the most likely
background sources. At the flux levels of the four sources under discussion,
the AGN counterparts have mean optical magnitude R$\sim 23-24$ mag. Although
counterparts at these magnitudes are normally trivially detected in the Subaru
imaging observations, we estimate that the surface brightness inside the
envelope of \object{KUG 0214-057} is high enough to preclude detection of
counterparts
fainter than $\sim 22$ mag. Thus only a small fraction of potential background
AGN counterparts would actually be detectable.

\subsection{X-ray properties of the ULX candidates}
The four ULX candidates discussed above are too faint for detailed X-ray
spectroscopy, but their X-ray colours can nevertheless provide some
constraints on the continuum shape and low energy absorption. Fig.3 shows
the
X-ray colours for the ULX candidates and, for comparison, the distribution of
hardness values for other X-ray sources in the SXDS which is of course
dominated by AGN in the field. Also shown is the position of E, the
source identified with a luminous AGN. Overall the four ULX candidates have
slightly higher values of HR1 than the average of all sources in the SXDS
field and somewhat lower values of HR3 than the average.\footnote{The location of
these sources in the X-ray hardness ratio plot (Fig.3) demonstrates that the nominal
spectrum used to compute the fluxes and luminosities (Table 1) is a reasonable
assumption given the quality of the data. The derived fluxes and luminosities
are of course sensitive to the spectral assumptions: increasing the assumed
column density to $N_H \sim 10^{21} \rm~cm^{-2}$ would increase them by a
factor of $\sim 1.5$, whilst the possible range of spectral indices also
introduces similar changes in the derived values, with steeper
(flatter) spectra
reducing (increasing) the flux and luminosity estimates by factors $\sim
1.5$--2.}
However the 
suggested power-law indices of $\Gamma \sim 2$ and low-to-moderate columns
($\sim 10^{21} \rm~cm^{-2}$) of sources C, A and B are quite
typical for accretion-powered ULXs (e.g. Roberts et al. \cite{rob04}, 
Colbert et al. \cite{col04}, Swartz et al. \cite{swa04}). Only source D appears unusual, with a
suggestion of a higher column and softer power-law index (Fig. 3 left panel).  
Since this source lies away from the main galaxy disc, and has a potential red
counterpart, this may indicate that it is a background source, though an
intriguing possibility is that its X-ray colours could be explained by a
``buried'' supernova remnant (cf. source CXOU J123029.5+413927 in NGC 4490,
Roberts et al. \cite{rob02}).

\begin{figure*}
\centering
\includegraphics[width=16cm]{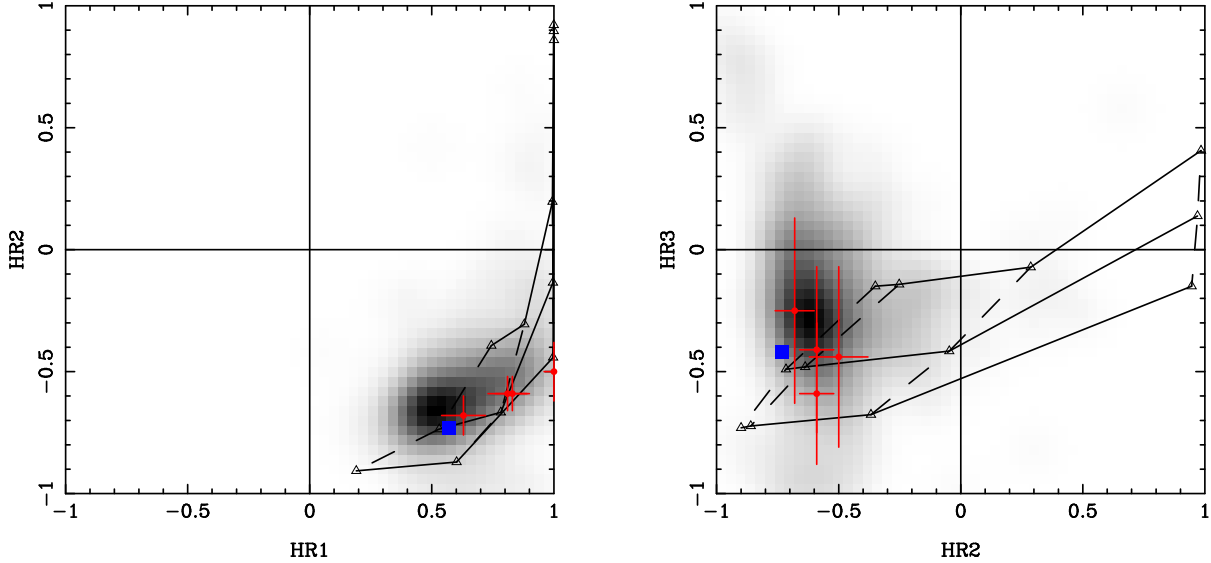}
\caption{X-ray colour-colour plots for the five X-ray sources: the
four ULX candidates (red points and error bars) and the source
identified with an AGN (E; blue square, error bars are $\sim 2$ times smaller than the
symbol plotted). The background grey-scale
shows the density of colour values for $\sim 500$ SXDS X-ray sources
with relatively low errors in the X-ray colour. {\sl Left:} HR2 vs
HR1; {\sl right:} HR3 vs HR2.  The HR values are defined as follows:
HR1 = (B-A)/(A+B); HR2 = (C-B)/(B+C); HR3 = (D-C)/(C+D) where A =
$C_X$(0.3-0.5 keV); B = $C_X$(0.5-2 keV); C = $C_X$(2-4.5 keV) and D =
$C_X$(4.5-10 keV) and the $C_X$ values are the vignetting-corrected EPIC
count rates. The grid overlaid on each panel corresponds to the
hardness ratios for power law spectra with indexes $\Gamma = [1, 2,
3]$ (top to bottom) and $N_H = [10^{20}, 10^{21}, 10^{22},
10^{23}]\rm\ cm^{-2}$ (left to right).}
\end{figure*}

A second X-ray diagnostic of the nature of these sources is temporal
variability. This is a particularly important test for relatively distant
sources such as these ULX candidates, where the spatial resolution of {\it
XMM-Newton\/} corresponds to physical scales of $\sim 2$ kpc, hence a
detection of variability (particularly on short timescales) strongly favours
an origin for the X-ray emission in a single accreting source as opposed to an
unresolved complex of many sources.  However, the limited nature of the raw
data (100 -- 200 counts per ULX candidate from all three EPIC detectors
combined) precludes all but the simplest of variability tests.  Comparative
photometry of the four ULX candidates, after splitting the observation into
three equal-length intervals ($\sim 15.9$ ks apiece), revealed no significant
variability.  A Kolmogorov-Smirnov (K-S) test against the hypothesis of a
constant source flux for the time series data divided into 5 ks intervals
indicated no significant variability in three of the sources, but source
B was found to have a $\sim 98\%$ probability of being variable ($\sim
2.4 \sigma$ significance). Hence at least one of the ULX candidates shows some
temporal evidence of being a single, accretion-powered source.

The X-ray properties of sources A \& C and in particular B appear
consistent with known ULXs.  Source D has slightly unusual X-ray colours,
perhaps consistent with a buried young supernova remnant, which would make
it a very rare source - only $\sim 10$ such sources with peak X-ray
luminosities in the ULX regime have ever been observed (Immler \& Lewin \cite{imm03}; 
this rarity also argues that the chances of the other ULX candidates 
also being such sources is very low indeed). There is of course still a
possibility that some of the candidate ULXs are no more than a
superposition of less-luminous sources clustered within the $\sim 5$
arcsec diameter XMM-Newton beam ($\equiv 1.8$ kpc at the distance of KUG
0214-057).  However, this would require a remarkable density of X-ray
sources - for illustration, the surface luminosity density of the 43 detected X-ray
sources in the central 1-arcmin ($\equiv 5.5$ kpc diameter) region of the
Antennae, calculated from the catalogue of Zezas et al. \cite{zez02} expressed as $\Sigma$
L$_{\rm X}$/area, is $\sim 9.5 \times 10^{38} ~\rm erg~s^{-1}~kpc^{-2}$.
In KUG 0214-057 one would require a higher density of $1.6 - 3 \times
10^{39} ~\rm erg~s^{-1}~kpc^{-2}$ for any of our candidate ULXs to be a
superposition of many fainter sources.  Hence the most likely
interpretation must be that we are viewing individual, accretion-powered
ULXs.

\subsection{Optical spectroscopy}
An optical spectrum of the brightest optical feature lying within the X-ray
error-circle for source B (second top and bottom panels in Fig.2) was obtained 
with the Subaru FOCAS instrument
in clear conditions with 1 arcsec. seeing on 25/12/2003 in an 
observation lasting a total of 1 hour (Akiyama et al. 2005, in preparation). 
The observation employed the  150mm$^{-1}$ low-resolution grating
and the SY47 filter to minimise order overlap,
providing wavelength coverage from 4700 to 9400\AA\ with
resolving power is R=250.
The resultant reduced
spectrum is shown in Fig.4. The line ratios for [SII]/H$\alpha$ and
[OIII]/H$\beta$ demonstrate that this is the spectrum of an HII region
rather than an AGN, albeit one with relatively high excitation when compared
with other Galactic and extragalactic examples (eg. Kennicutt et al. \cite{ken00}).
In fact the observed ratios are close to HII/AGN dividing line, indicating
that a hard X-ray photoionizing source may be located in the nebula as
expected for a ULX (eg. Pakull \& Mirioni \cite{pak03}, Roberts et al. \cite{rob03})
although it is doubtful that source B could provide all
the ionising flux.

\begin{figure}
\centering
\includegraphics[width=8.5cm]{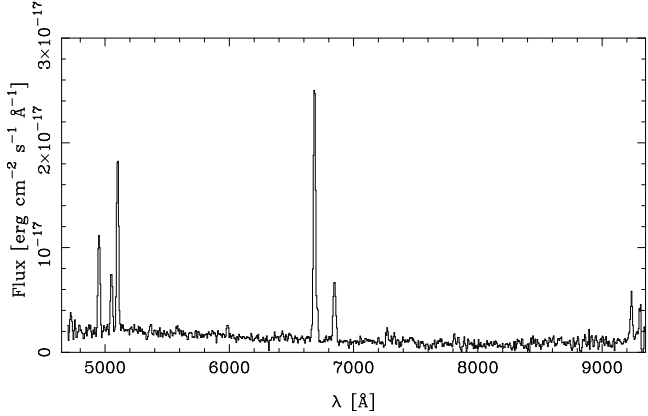}
\caption{FOCAS optical spectrum of brightest knot in nebulosity within
 error-circle of source B (see Fig.2). The bright emission lines are
  identified as H$\alpha$ (+ [NII]), [OIII] $\lambda 4959, 5007$\AA\ ,
H$\beta$, [SII] $\lambda 6716, 6731$\AA\ and [SIII] $\lambda 9069$\AA\ at z=0.018.}
\end{figure}
 
\subsection{Nature and properties of the galaxy \object{KUG 0214-057}}
From its optical appearance, \object{KUG 0214-057} is clearly a late-type galaxy with
very significant recent star formation. Its inclusion in the KUG catalogue 
of UV-excess galaxies reinforces this view as UV-excess is most commonly an
indicator of relatively strong star formation.
There is some indication of disruption
to the outer spiral arms. It is also interesting to note the small companion
galaxy (\object{APMUKS(BJ) B021449.37-054240.9}) $\sim 1.5$ arcmin. to the SW which may
be interacting with \object{KUG 0214-057}, and the very faint nebulosity lying a
similar distance to the SE (see Fig. 1c). This nebulosity might actually be a
fragment left over from an earlier interaction.

\object{KUG 0214-057} was not detected by IRAS and hence there is no mid-IR data to
constrain the star formation rate.  Similarly, there are no corresponding
measurements of the flux in H$\alpha$, UV or radio continuum, hence these
classic methods for deriving a star forming rate are also not available.
However, recent work has shown that the integrated X-ray luminosity of
luminous high mass X-ray binaries in star forming galaxies can also be a
measure of star formation rate (see e.g. Grimm, Gilfanov \& Sunyaev \cite{gri03}).
We therefore estimate an X-ray star forming rate in the following manner.  We
assume a starburst-like X-ray luminosity function slope of $\gamma = 0.5$
(Kilgard et al. \cite{kil02}) for sources in \object{KUG 0214-057}, and normalise this to the
integrated flux of the four detected sources ($\sim 2.8 \times 10^{40} \rm
~erg~s^{-1}$) over the $4 - 20 \times 10^{39} \rm ~erg~s^{-1}$ range (i.e.
between the minimum detected luminosity of the ULXs, and the apparent break
luminosity for galaxian luminosity functions noted by Grimm, Gilfanov and
Sunyaev \cite{gri03}).  We then integrate below this range (i.e. between $0 - 4 \times
10^{39} \rm ~erg~s^{-1}$) to estimate the ``missing'' flux from discrete
sources below our detection limit\footnote{We caution that the actual detection limit could be
lower, which would lower the estimated contribution from undetected sources,
for example if the detection limit is $2 \times 10^{39} \rm ~erg~s^{-1}$ then
the estimated missing flux reduces to $1.3 \times 10^{40} \rm ~erg~s^{-1}$
On the other hand, the PSF wings of the bright QSO obscure our view of the
southern regions of the galaxy, hence we potentially miss detecting sources in
this region which could lead to us underestimating the luminosity of the
galaxy. Indeed there is clear evidence that the profile of source E is
asymmetric (Fig.1), implying the presence of one or more further faint
objects in the vicinity of source E.},
which totals $2.3 \times 10^{40} \rm~erg~s^{-1}$.
Hence we estimate a total X-ray flux from luminous point-like X-ray
sources of $\sim 5 \times 10^{40} \rm ~erg~s^{-1}$.  From Fig. 7 of Grimm,
Gilfanov \& Sunyaev \cite{gri03} this converts to a reasonably high star formation
rate in the range of $\sim 3.5 - 6$ M$_{\odot}$ yr$^{-1}$.

To put this star formation rate in perspective, we estimate the mass of 
\object{KUG 0214-057} using the stellar $M/L$ correlation coefficients of Bell and De Jong \cite{bel01}.  
In particular, we follow Colbert et al. \cite{col04} and use coefficients
from the formation with bursts model, and we use optical/IR magnitudes from
NED of $B = 15.26$ (Maddox et al. \cite{mad90}) and $K = 13.16$ (2MASS measurement).  
These convert to give $log_{10} L_K \approx 42.3 \rm ~erg~s^{-1}$, $log_{10}
(M/L_K) \approx 0.25$ and hence $M \approx 1.2 \times 10^{10} M_{\odot}$.  
Finally, by normalising the star formation rate to the galaxy mass we find
that \object{KUG 0214-057} has a rate of $\sim 4$ (in units of $10^{-10} \rm yr^{-1}$),
which from Table 1 of Grimm, Gilfanov \& Sunyaev \cite{gri03} is very similar to the
rate per unit mass in the archetypal local starburst M82 ($\sim 3.6$).  
However, in contrast to M82, the locations of both the ULXs and optical knots
demonstrate that the star formation activity appears to be prevalent 
throughout the spiral arms of \object{KUG 0214-057}, rather than dominant in the
nucleus.  This is another indicator that the high star formation rate is due
to an interaction: similar star formation morphology and ULX locations are
seen in more nearby interactions, e.g. in NGC 4485/90 (Roberts et al. \cite{rob02})
and M51 (Terashima \& Wilson \cite{ter04}). However, we note that both the star
formation rates and the integrated and peak luminosities of the X-ray source
populations are far higher in \object{KUG 0214-057}.  Indeed, Ptak \& Colbert \cite{pta04}
estimate that only $\sim 7\%$ of galaxies possess one or more ULXs with
L$_{\rm X} > 2 \times 10^{39} \rm ~erg~s^{-1}$; \object{KUG 0214-057} potentially has
four.  This high incidence of ULXs is only bettered by the most intense local
starbursts, usually in galaxy collisions e.g. NGC 3256 (Lira et al. \cite{lir02}), the
Cartwheel (Gao et al. \cite{gao03}).

\section{Conclusions}
We have shown that the galaxy \object{KUG 0214-057} probably contains up to four ULXs
with luminosities at the high end for such objects. Given the low mass
of inferred for this galaxy, the high star formation implied by the presence
of these ULXs indicates that this previously relatively obscure galaxy
possesses remarkably intense starburst activity. These ULXs, discovered
in the SXDS, are drawn from an unbiased extragalactic survey rather than
pointed observations of specific galaxies, allowing a estimate to be made
(albeit very crude) of the contribution ULXs make to the overall
extragalactic source population. The three or four sources reported here
comprise $\sim 0.5$\% of the total source population detected in the SXDS.
Several other ULX candidates exist in the SXDS (these have yet to be fully
explored), demonstrating that this may be an underestimate 
of their total contribution. ULXs are thus a minority but not negligible
population in deep surveys.

\begin{acknowledgements}
This work is based in part on data collected at Subaru Telescope, which is operated 
by the National Astronomical Observatory of Japan, financially supported in part by 
a Grant-in-Aid for Scientific Research (No. 15740126) by the Japanese Ministry of Education, 
Culture, Sports, Science and Technology. 
\end{acknowledgements}

\end{document}